\documentclass[10pt]{IETBook}
\usepackage{booktabs}
\usepackage{tabularx}
\usepackage{float}
\usepackage{array} 

\begin{document}

\rhbooktitle{Emerging Trends in Multimedia Data Security and Computational Intelligence}

\markboth{Emerging Trends in Multimedia Data Security and Computational Intelligence}{Emerging Trends in Multimedia Data Security and Computational Intelligence}

\cauthor{Hamish Alsop\thanks{School of Computing, Engineering and the Build Environment, Edinburgh Napier University, UK},
Leandros Maglaras\thanks{School of Computer Science and Informatics, De Montfort University, Leicester, UK},
Helge Janicke\thanks{Centre for Securing Digital Futures, Edith Cowan University, Perth, WA 6027,
Australia},
Iqbal H. Sarker\thanks{Centre for Securing Digital Futures, Edith Cowan University, Perth, WA 6027,
Australia}, and
Mohamed Amine Ferrag \thanks{Department of Computer and Network Engineering, United Arab Emirates University, UAE}}

\chapter{Innovating Augmented Reality Security: Recent E2E Encryption Approaches}

End-to-end encryption (E2EE) has emerged as a fundamental element of modern digital communication, protecting data from unauthorized access during transmission. By design, E2EE ensures that only the intended recipient can decrypt the information, making it inaccessible even to service providers. Yet, this powerful safeguard of individual privacy and digital trust also introduces a paradox: it can simultaneously prevent law enforcement efforts by hiding potential malicious activities. This paper examines the dual role of E2EE, its critical importance to privacy, the challenges it
introduces, and highlights recent innovative approaches developed thought the use of AR technology.
\index{fa series index text index!bcde index text index text index text!cdef index text index text} 
\index{fb series index text index text index!bcde index text index text!cdef index text}\index{fb series index text index text index!cdef index text index text index text}

\section{Introduction}

The application of end-to-end encryption (E2EE) is a critical component of mobile network security and privacy preservation of social media content. E2EE is based on various cryptographic techniques, including symmetric and asymmetric encryption, and blockchain mechanisms.  These techniques are systematically used to enhance the security of social media applications and other digital platforms \cite{pansy2025end}. The primary goal of implementing encryption methods is to raise the overall security level, ensure user privacy, and protect user data against unauthorized access to sensitive information. By integrating E2EE, social media platforms aim to secure the vast digital landscape where users share their thoughts, ideas, and experiences, reinforcing trust and enabling global connections in a more secure environment.

E2EE is a cornerstone of modern digital security, providing confidentiality, integrity, and authenticity for communications \cite{maglaras2025end}. It ensures that only the sender and the intended recipient can read messages, with no intermediaries (including service providers or even the platform itself) able to access the plaintext content. This is achieved by encrypting the data on the sender's device and decrypting it only on the recipient's device. However, the effectiveness and widespread adoption of E2EE are influenced both by its inherent technological complexities and by the crucial role of human factors.

Recently, the FBI and the Cybersecurity and Infrastructure Security Agency (CISA) issued a warning to smartphone users about the risks of sending unencrypted text messages between iPhones and Android devices \cite{FBI2024}. They noted that people share too much information through text messages, from casual conversations to sensitive data such as passwords and usernames. This presents a serious problem, as this kind of data is exposed and vulnerable to theft. The potential for cybercriminals to intercept these unencrypted messages makes the security risk critical. These messages are vulnerable to interception by cybercriminals, which could expose sensitive information. This highlights the importance of using E2EE-enabled applications to protect your data from potential threats.

As concerns about surveillance, data breaches, and the misuse of personal information continue to grow, there has been a fundamental shift in the way users approach and deal with digital security. The traditional model, which relied on "trusting the provider" to secure data on their servers, is being replaced by a model that prioritizes cryptographic assurance \cite{goel2024improving}. This shift reflects a recognition that even a trustworthy service provider can be a target for attackers or may be forced by criminals or even legal entities to hand over user data. By shifting trust from a centralized entity to an ''unbreakable'' encryption mechanism, users can gain greater control and confidence in the privacy and security of their communications. The same pattern is observed in big companies too, which are shifting from public cloud to private cloud, a trend known as cloud repatriation. This shift is driven mainly by security concerns and the need for greater control over infrastructure and data \cite{comapniespricloud}. 

This evolution is a direct result of the increasing awareness that personal data is a valuable asset that needs to be protected from both malicious actors and well-intentioned but potentially unsecure service providers \cite{birch2023data}. The move towards "trusting the cryptography" is a powerful driver for the adoption of E2EE which minimizes the risk of data exposure at every point in the transmission chain, from the moment a message is sent to the moment it is received, thereby providing a robust defense against modern digital threats.

While E2EE protects data in transit, a significant security gap remains: endpoint vulnerabilities \cite{isobe2021security}. Traditional security models often fail to address threats like malware and spyware that can compromise end devices, especially mobile phones. Even with end-to-end encrypted services like Signal, where data isn't stored in a cloud backup, messages are still created and consumed as clear text on the user's device. This makes the data susceptible to attackers who can access a compromised device, bypassing the encryption mechanism. This critical weakness has become a major focus of ongoing research and development \cite{fabrega2024injection}. To provide comprehensive protection, security solutions must extend beyond the network to secure the endpoints where data is first created and last accessed. The goal is to develop new methods that can protect data in its unencrypted state, guarding against threats that have already breached the device's defenses.

This chapter explores the current landscape of end-to-end encryption (E2EE) solutions, focusing on their role in securing digital communication and protecting user privacy. 
The main areas covered are:

\begin{itemize}

\item Privacy Considerations: We examine various privacy aspects, including location, physical and mental state, social life, and media.
\item  Emerging Threats: We analyze new threats, particularly those related to insecure third-party applications, the human factor in cybersecurity attacks, and client-side scanning (CSS) solutions.
\item Novel Solutions: We present state-of-the-art E2EE solutions, including those that integrate with augmented reality and encrypted keyboard technologies.
    
\end{itemize}

By exploring the vulnerabilities in existing systems and the human element in cybersecurity attacks, this work highlights the need for innovative frameworks. The goal is to contribute to the ongoing conversation about how to effectively balance privacy, security, and ethical considerations with technological advancements in the digital world.

\section{Technological Perspectives}
End-to-end encryption (E2EE) represents a paradigm shift in digital security, moving the focus of trust from centralized service providers to the user's own device. By ensuring that data is encrypted on the sender's device and can only be decrypted by the intended recipient, E2EE provides an high level of data confidentiality and privacy. This section focuses on the core cryptographic mechanisms used in E2EE while it briefly presents some challenges of this technology.

From a technological perspective, E2EE's strength lies in a sophisticated combination of cryptographic mechanisms. The foundation is built upon Asymmetric (Public-Key) Cryptography, which utilizes a pair of mathematically linked keys: a public key and a private key. A user's public key can be freely shared, as it is used solely for encrypting data. The private key is kept secret and is the only key capable of decrypting data encrypted with the corresponding public key. This two-key system is also used for Digital Signatures, where a message is "signed" with a sender's private key, allowing the recipient to verify the sender's authenticity and the message's integrity using the sender's public key. While asymmetric cryptography is excellent for secure key exchange, its computational demands make it inefficient for encrypting large data streams \cite{kshetri2024algotric}.

This is where Symmetric Cryptography comes into play. It relies on a single shared secret key for both encryption and decryption, making it significantly faster and more efficient for bulk data transmission. The common practice in modern E2EE systems, such as the Signal Protocol, is to create a Hybrid Encryption System. Here, asymmetric cryptography is used to securely exchange a one-time session key. This session key is then used with a symmetric algorithm to encrypt the actual message content. This approach leverages the security of asymmetric cryptography for key exchange and the efficiency of symmetric cryptography for data transmission \cite{zhang2021overview}.

A crucial feature of modern E2EE protocols is Perfect Forward Secrecy (PFS). This principle ensures that if a long-term private key is compromised, it does not compromise past or future session keys. Each new communication session generates a new, ephemeral session key, so a breach of one key does not expose past conversations. This is a critical element for long-term privacy and security. Authentication and Integrity are also vital, employing cryptographic hashes and digital signatures to verify the sender's identity and confirm that the message has not been tampered with in transit \cite{adebayo2025end}.

E2EE comes with some challenges and open issues. One example is the lack of a universal Interoperability and Standardization across different platforms which is a significant impediment. The fragmentation of E2EE protocols can create security gaps and prevent seamless secure communication across different services. For enterprises, E2EE introduces Blind Spots for Enterprise Security \cite{baker2025tunnel}. Finally, the very nature of E2EE, which decentralizes trust, means there is an Absence of a Central Recovery Mechanism, which, while a security feature, poses a practical challenge for users who need to recover their data. The inherent trade-off between absolute privacy and the practical need for data recovery and enterprise-level control is a continuous point of friction. All this open issues, challenges and considerations are further presented and analyzed in the upcoming sections.

\section{Privacy Considerations}

The widespread adoption of E2EE is driven by several key privacy advantages:

{\bf Confidentiality} At its core, E2EE guarantees confidentiality by preventing unauthorized access to user content of messages and data. Unlike traditional encryption methods where data may be decrypted by the service provider, E2EE ensures that the plaintext remains visible only to the communicating endpoints. This is a fundamental safeguard for sensitive personal, financial, and professional information.

{\bf Protection from Intermediaries}  E2EE fundamentally shifts the trust model away from the service provider. In an E2EE system, the provider acts as a simple channel for transmitting encrypted data, unable to read or exploit the content of the communications. This prevents the broad collection and aggregation of user data by companies for commercial purposes, such as targeted advertising, and protects against potential misuse of user data by the service provider itself \cite{heldt2022eu}.

{\bf Safeguarding Against Data Breaches} One of the most significant threats in the digital age is the compromise of the server. When a service provider’s servers are breached, the unencrypted data stored on them can be stolen. With E2EE, even if a server is compromised, the encrypted data remains unreadable to the attackers, mitigating the impact of the breach on user privacy \cite{jamal2024safeguarding}. Attackers may gain access to encrypted messages, but without the private keys from the users' devices, the content remains secure.

{\bf Prevention of Mass Surveillance} E2EE serves as a powerful deterrent to massive monitoring by state and non-state actors \cite{cao2022mass}. By making it technically infeasible for third parties to passively collect and analyze communication content on a large scale, E2EE systems prevent mass surveillance efforts, thereby protecting freedom of expression and association.

\subsection{Limitations and differences in E2EE Privacy}

While E2EE provides strong guarantees for data confidentiality, its effectiveness is limited by several factors:

{\bf Metadata Exposure} A critical and often overlooked vulnerability in E2EE systems is the exposure of metadata. Although the content of communications is encrypted, metadata such as the sender and recipient, the time and frequency of communication, device information, and IP addresses, is often transmitted in the clear \cite{khader2023assessing}. Analysis of these metadata can reveal significant insights into an individual's social network (social graph), daily routines, and associations, even without ever viewing the message content. For example, traffic analysis can deduce communication patterns and locations. Although some privacy-conscious providers, such as Signal, have made efforts to minimize metadata collection, it remains a persistent challenge.

{\bf Endpoint vulnerability} The security of an E2EE system is only as strong as its weakest link, which is often the endpoints, thee user's device. Malware, spyware, and other forms of malicious software can completely bypass E2EE by infecting a device and capturing data before it is encrypted or after it has been decrypted. Similarly, physical access to a device can allow an attacker to read messages directly \cite{maglaras2025end}. 

{\bf Insecure Cloud Backups} A common practice that undermines E2EE is the use of unencrypted cloud backups. Services such as WhatsApp, for instance, offer users the option to back up their chats to platforms like Google Drive or iCloud. By default, these backups are not encrypted, making the data vulnerable to access by the cloud provider or through a compromise of the user’s cloud account. Although some providers are now introducing features for encrypted backups, the user often has the option to enable these security measures.

{\bf Third-Party Integration} E2EE privacy guarantees can be lost when communications leave the secure environment of the messaging app. This is particularly relevant in business contexts. When, for example, a customer service conversation is routed through an E2EE platform, the moment it is integrated into a third-party Customer Relationship Management (CRM) system or other external database, the data becomes unencrypted and accessible to the business owner and its partners.

{\bf The "Backdoor" Debate} The fundamental conflict between privacy and "lawful access" continues to be a central debate. Governments and law enforcement agencies often argue for the inclusion of "backdoors" or other mechanisms that would allow them to bypass E2EE to access criminal communications. However, this proposal is fundamentally at odds with the cryptographic principles of E2EE, since a backdoor for one party is a vulnerability for all and could be exploited by malicious actors \cite{scott2024technical}.

{\bf AI models and Privacy} The integration of artificial intelligence (AI) features, such as smart replies, content summarization, and predictive text, presents a new and complex privacy challenge. In order to perform these functions, AI models typically require access to plaintext data. This creates a dilemma: Should the data be sent to the cloud for processing, where it loses its E2EE protection, or should the AI processing be done on the user's device, which can be computationally intensive? The concept of "Private Cloud Compute" is emerging as a potential solution, where encrypted data is processed in a secure enclave without being decrypted \cite{ngo2018cloud}.

\section{Threats and Challenges to E2EE's Integrity} 
While E2EE provides a high level of security, it is not without its threats and challenges. This section presents all current threats to E2EE solutions.

\subsection{Direct Threats to E2EE Security}
E2EE integrity can be compromised by a number of direct threats that exploit weaknesses outside of the encryption algorithm itself. These threats often target the devices at either end of the communication or the communication process before or after the encryption takes place.

\subsubsection{Endpoint Compromise}
As useful as E2EE is at protecting data in transmission, cannot protect that data if the endpoint itself is compromised. If malware such as Spyware or Stalkerware are present and running on a user's device, then this can essentially bypass the functionality of E2EE by capturing data such as keystrokes or screen recording, allowing the attacker to see the data in plain text either before the data is encrypted, or after it has been decrypted. A famous example of this is the Pegasus spyware kit, which has been used by advanced attackers to infect smartphones and then effectively bypass E2EE by reading messages directly from the device's memory, essentially making the encryption "meaningless" once installed ~\cite{pegasus}. 

Attackers can also manage to circumvent E2EE by exploiting physical access to a device. If an attacker is able to steal or unlock a device, they will be able to read any stored messages or even impersonate the real user, although this risk can be reduced with the use of strong passwords or disk encryption. 

The risk of malware can also be reduced by ensuring strong endpoint security, this can be achieved by keeping devices up to date; both in terms of OS version as well as regular security patches and application updates. These risks can be even further mitigated by deploying anti-virus or anti-malware software on the device~\cite{ibmE2EE}. 

Even the more advanced countermeasures, such as isolating encryption keys within secure hardware locations, are not immune to threat. Some sophisticated types of malware may exploit zero-day vulnerabilities or abuse OS privileges in order to extract keys from the trusted execution environments. While these hardware protections are effective and can reduce the attack surface, they are not completely bulletproof. 

Having strong endpoint security is clearly an advantage as it can help to protect and preserve the confidentiality of messages, even while under attack, however this does somewhat create an environment of complete trust between the user and their device, which can be taken advantage of. 

An open research problem is malware detection on resource constrained or wearable devices and ensuring E2EE apps can run on a "clean" platform. Recent prototypes such as ARSecure (see section 1.6.1) even propose offloading encryption to a secondary, secure device (AR Glasses) in order to keep the whole process separate from a potentially compromised phone~\cite{alsop2024arsecure}.

\subsubsection{Client-Side Scanning (CSS)}

Client-Side Scanning (CSS) refers to the practice of analysing content on a user's device either before or during the encryption process. This is typically implemented to detect illegal material—most notably child sexual abuse material (CSAM), prior to it being transmitted through an encrypted channel~\cite{eff_css1}. Proponents argue that CSS enables content moderation and law enforcement interventions without technically “breaking” the encryption, since technically the ciphertext is still transmitted across the network. In reality however, CSS completely undermines the core privacy guarantee of E2EE by performing surveillance at the point of message creation. It can be thought of as someone reading and checking your letters before they get sealed in the envelope~\cite{isoc_css}. 

Whether it is mandated by government or implemented voluntarily by platform providers, CSS software introduces new attack surfaces and effectively functions as a backdoor that can be easily expanded beyond it's initial purpose~\cite{eff_css2}. Critics argue that allowing any form of client-side inspection fundamentally compromises the confidentiality principle promised by E2EE: that only the sender and recipient can access message contents~\cite{eff_css3}. The introduction of CSS software can become a slippery slope, as once implemented, the list of “prohibited content” can be expanded silently, making CSS vulnerable to abuse. Topics such as political opinions, dissent, or otherwise lawful content could be added to matching databases without public knowledge~\cite{eff_css4}. 

Embedding these kinds of scanning tools into devices opens up a whole host of security risks, as the tool itself could harbour vulnerabilities that allow it to be easily repurposed by attackers, and therefore putting many more users at risk.
Implementing CSS can have a major impact on user trust levels, effectively eroding a user's confidence that the conversations that they are having are, in fact, private, completely contradicting the privacy-by-design philosophy of E2EE ~\cite{isoc_privacy_risks}. 

Alternative approaches (discussed later, e.g. homomorphic encryption in 1.6.3) are being researched in order to bring about some form of content checking without actually exposing plaintext, but these technologies are in early stages and are often impractical in real world use~\cite{isoc_he}. Furthermore, unless users have the ability to disable CSS or choose apps that do not implement it, they may be forced to use platforms that violate their privacy. If this kind of legislation comes into place in the future, this freedom of choice may be further diminished. 

Basically, while there is the argument that CSS aims to prevent the spread of illegal content and can be viewed as a vital tool to protect people from exposure to harmful content, it does so at such a substantial cost to both the privacy and security of the platform's users. This debate still remains a hot topic of conversation as researchers and policymakers continue to question whether a secure CSS implementation is even feasible, and more more importantly, who gets to decide what content is deemed “illicit”~\cite{alsop2024arsecure}. This is especially true given the current and proposed legislation that may come into play later in 2025.

\subsubsection{Key Management Vulnerabilities}

The strength of E2EE rests on the secrecy and authenticity of its cryptographic keys, however, improper key management can lead to the introduction of a whole host of vulnerabilities, even if the encryption algorithms are strong~\cite{kiteworks_keys}.

These key management issues can include:
\begin{itemize}
    \item Weak key generation (e.g. using low quality random number generators or short keys)
    \item Improper key storage (keys left unencrypted on disk or in cloud backups)
    \item Insecure key backup mechanisms
    \item Lack of key rotation
    \item Failure to revoke/refresh keys when needed
\end{itemize}

For example, if a messaging application backs up their E2EE keys to the cloud in plaintext for "user convenience", an attacker who gains access to that cloud storage would be able to obtain the keys and therefore be able to decrypt conversations, completely defeating E2EE entirely. Similarly, if an application allows users to reuse the same long-term keys indefinitely, then a single compromise could potentially expose all past and future messages.

Key authenticity is another aspect that needs to be considered. This can be exploited by an attacker by tricking a user into accepting a fraudulent public key (perhaps by exploiting a flawed key exchange), and use this to perform a man-in-the-middle (MITM) attack, allowing them to read all communications despite the user believing that E2EE is in place.

Strong key management techniques such as using vetted protocols that use ephemeral sessions keys with forward secrecy \cite{omora2022signal}, or storing long term keys in secure keystores, and providing secure mechanisms for users to verify keys are all good methods of mitigating the threats that can come with poor key management. 

\subsubsection{Implementation Flaws and Protocol Vulnerabilities}

Even if all the theory behind the cryptography is technically sound, things such as software bugs or design flaws in the application implementation can break the E2EE. In the real world, encryption protocols are often complex, and mistakes such as improper random number usage, or buffer overflow errors, or logic issues can create attack surfaces for attackers to exploit and break the encryption. A recent example of this was highlighted back in 2021 when security analysis of the E2EE protocol used by the application "Zoom" revealed several vulnerabilities which allowed insider attackers to impersonate users in online meetings~\cite{barrera2021zoom}.

Over the years we have seen earlier versions of other messaging applications fall victim to issues like Downgrade Attacks (where the attacker tricks the app into using a weaker cypher or an out of date protocol), Malleability Attacks (where the cyphertext is actually altered), or side-channel leaks.

\subsubsection{Social Engineering and Human Error}

Even when applications have strong cryptography in place, the human element still poses a major threat to E2EE in the form of social engineering and general human error. Phishing attacks can be used by threat actors to trick users into revealing authentication codes or clicking malicious links and potentially bypassing the encryption altogether by targeting the human, which in most cases, is the weakest link \cite{informationweek2022e2ee}. Since E2EE doesn't verify identity by default, there are many applications that rely on trust-on-first-use (TOFU) key exchanges, effectively leaving users vulnerable to impersonation or even man-in-the-middle (MITM) attacks if they fail to verify the keys manually \cite{signalTOFU}. Popular messaging applications such as WhatsApp and Signal do offer safety number or QR code verification, however we have seen that it is very common for users to actually skip these steps, weakening the system's trust guarantees \cite{olson2020usability}.

Strong identity verification is further complicated by the tension between privacy and authentication. Platforms could reduce the risk of impersonation by enforcing strict ID checks but this would in turn undermine anonymity, which is something that is valued by journalists, whistleblowers, and activists, who rely on the security and secrecy of E2EE messaging platforms \cite{yeo2025}. 

We can see these risks in real life examples, such as in 2025, when a U.S. government Signal group was unintentionally compromised by a contact being added without the proper verification \cite{rocketchat2025signalbreach}. In the end, E2EE can protect data in transit but it cannot protect against the user being tricked or careless. In order to ensure meaningful confidentiality, the platform requires more than just the encryption alone, but also a sturdy identity assurance method as well as secure user practices \cite{eff2023usererror}.

\subsubsection{Government Pressure and Legal Mandates}

Governments all over the world have leveraged legal mandates and/or technical coercion in order to compel decryption, or the insertion of backdoors into encrypted communication systems. This is a practice that completely undermines the privacy guarantees of end-to-end encryption. 
Examples of this can be seen in the UK, where the Investigatory Powers Act (IPA) has been wielded in the past to issue "Technical Capability Notices", which legally obligate providers, such as Apple, to disable or weaken encryption features. In response to this Apple actually withdraw its Advanced Data Protection (ADP) service from the UK market rather than build surveillance backdoors \cite{apnews2025applebackdoor}.

This "Going Dark" concern (law enforcement's inability to access encrypted communications) is historically invoked to justify these measures, however there is little evidence to suggest that mandated backdoors actually improve investigative outcomes \cite{arxiv2021backdoors}. Also, adding in deliberately engineered vulnerabilities, even when they are implemented with strict oversight, will inevitably introduce systemic weaknesses that are inherently exploitable as well as being difficult to contain \cite{betanews2025backdoordebate,thirdway2020weakenedencryption}.

Then there are governance mechanisms such as key disclosure laws which add further complications to the situation. In many jurisdictions, such as the UK, Australia, or India, individuals or service providers may be required by law to surrender keys or decrypt data under warrant or threat of imprisonment, which then even further erodes E2EE's efficacy as well as the autonomy of the users \cite{wiki_keydisclosure,ripal2000}.

More recently, the European Union has advanced the so-called "Chat Control" legislation, that was formerly known as the Child Sexual Abuse Regulation (CSAR). This new proposal would mandate that client-side scanning of communications be put in place before encryption, which forces E2EE services like Signal, or WhatsApp, or Telegram to implement surveillance mechanisms into their apps. Many would argue that putting such measures in places essentially defeats the purpose of using E2EE communications and creates an infrastructure for mass monitoring \cite{chatcontrol2025techradar, chatcontrol2025blackmail}.

\subsubsection{Compromised Trust in Software Providers} 

Trusting software providers as the custodians of end-to-end encryption can prove dangerous when those providers are infiltrated themselves, or worse, when their updates are weaponised. The 2020 SolarWinds supply chain compromise, where attackers embedded their SUNBURST backdoor into a legitimate Orion software update, shows us that even the officially signed updates can be altered at the source and can potentially expose thousands of private and public sector organisations to all kinds of threats \cite{wired_solarwinds_supplychain, cisa_solarwinds_alert}.

Similarly, the NotPetya outbreak back in 2017 emerged from a compromised update of Ukraine's accounting software called M.E.Doc. This update pushed out destructive malware under the guise of legitimate patches and ended up causing billions in global damages \cite{wired_notpetya, axios_supplychain2017}.

Beyond platform-level breaches however, lies another threat, the presence of malicious 3rd party applications within mobile ecosystems. These applications continue to broadcast or leak cryptographic secrets such as key material, effectively undermining the assumption that installed applications respect the user's privacy \cite{wired_hacker_lexicon_supplychain}.

These examples show us a simple truth: the integrity of E2EE relies not just on the cryptographic protocols, but also on the security of the entire software delivery chain. Some effective countermeasures include things such as reproducible builds, cryptographic validation of code, narrow and verifiable dependency chains, and aggressive vetting of all software components that enter client devices.

\subsubsection{User Experience (UX) Complexity}
The most secure end-to-end encryption mechanisms are worhtless if users cannot, or will not, use them. Historically E2EE tools have not been able to achieve broad adoption largely due to their poor usability.

Studies that have been done on email encryption systems such as PGP, S/MIME, and pEp have shown that over 60\% of users were completely unaware of their existence and struggled with the cumbersome management of public keys and complicated setup procedures. This is even though 78\% of participants said that encrypted email communication was important to them \cite{reuter_usability_email}.  

When it comes to messaging applications such as WhatsApp, Viber, Signal, and Telegram, user studies have seen both high false-accept and false-reject rates when it comes to remote key-fingerprint verification, especially in situations where the users cannot easily compare the codes face to face \cite{shirvanian_codeverif}. These findings make it clear that user convenience often beats out actual security, since users tend to favour workflows that "feel good enough", rather than those that are more secure but potentially unusable \cite{userlab_registration_vs_exchange}. 

It seems that designing E2EE messaging applications for mass adoption means decoupling the cryptographic complexity from the user interface. The UX can be simple and easy to use while still implementing strong encryption methods. The user interface should be intuitive, standardised, no hidden menus or "crypto jargon", and not rely on manual key exchanges. The best implementations must implement strong security while making it effectively invisible to the user. 

\subsubsection{Key Management at Scale}
When it comes to E2EE messaging systems, cryptographic key management is not a one-time task, it is very much a continuous, large scale operation. At the core, each user maintains a long term identity key pair, but modern protocols such as Signal's X3DH or Double Ratchet require the generation, storage, and rotation of vast numbers of short lived ephemeral keys - often one per session, per device, and in some designs, even one per message \cite{omora2022signal}. If you increase this up to a global scale, it translates to billions of keys being created, distributed, and retired each day across a diverse set of network conditions as well as device types.

This struggle is compounded by the presence of multi-device support. Services like WhatsApp and Signal have to be able to syncronise E2EE conversations across all of a user's devices without ever centralising the private keys on the provider's servers. This issue forces the development of intricate key-sync mechanisms which tend to involve E2EE key transfer sessions between devices \cite{whatsapp_multidevice}. 

As if that wasn't enough, group messaging adds yet another dimension to the already complex issue. Secure group chat protocols such as Sender Keys in Signal, must update and redistribute the group encryption keys any time membership of the group changes, this ensures both forward and backward secrecy, even in groups with hundreds of participants \cite{signal_senderkeys}.

Key loss and recovery remain somewhat unsolved at scale. If a user reinstalls the application or loses the device, and the provider does not store the private keys, then their conversations cannot be decrypted. Some platforms mitigate this with the use of encrypted backups, but these then re-introduce server side trust assumptions, and therefore create new attack surfaces for threat actors to exploit \cite{ibmE2EE}. Being able to balance strong security guarantees with practical recovery options is likely one of the most challenging operations problems within E2EE messaging.

In order for E2EE messaging applications to be successful at scale, they must be able to solve the issues in four competing demands: strong cryptographic assurances, low latency in messaging, resilience to device churn, and minimal user friction. If the provider fails in any of these areas then it will either degrade the user experience, weaken the E2EE model through the use of insecure workarounds, or even drive users to use a less secure platform.

\subsubsection{Performance and Resource Overhead}
End-to-end encryption adds meaningful overhead to messaging on mobile devices, in both CPU usage as well as battery consumption. Cryptographic operations like key agreement, symmetric encryption, and ratchet advancement can significantly tax constrained hardware, potentially draining batteries or causing noticeable lag \cite{annablackwell_performance}. 

This kind of performance degradation becomes particularly acute in low-bandwidth or high-latency environments where prolonged encryption or delays in handshakes not only slow down messaging but can completely interrupt user interactions. Studies of Android messaging apps (WhatsApp vs. IMO) demonstrated that the app design and optimised code paths can result in noticeably different CPU utilisation during VoIP sessions, making it clear that encryption strategy as well as implementation matter deeply in terms of real-world performance \cite{olivwhatsapp_performance}.

\subsubsection{Interoperability Issues}
Interoperability between E2EE messaging applications is not a luxury, but rather it seems to be more of a legal mandate, however actually managing to integrate platforms such as WhatsApp or iMessage into a seamless cross-platform ecosystem, unsurprisingly, introduces severe security and architectural complexity. As the EU's Digital Markets Act (DMA) demands cross-app messaging, it forces developers to try and resolve divergences in encryption, identity, and metadata protocols that where never designed or intended to interoperate \cite{blessing_oneprotocol}. 

Even in cases where the services adopt the same underlying protocol (e.g. Signal protocol), differences in implementation, session storage, key derivation, and backup methods can ruin or degrade the security if it is not carefully aligned. The concept of interoperability is not impossible though, WhatsApp was able to achieve this, however their adoption of third-party interoperability required novel mechanisms in order to preserve E2EE while allowing the cross-application connectivity. While this shows that it is possible, such engineering feats are highly bespoke and often fraught with risks \cite{fb_interop}.

\subsubsection{Regulatory Compliance and Enterprise Management}

For E2EE messaging platforms that are operating in regulated industries, the encryption strength alone does not ensure lawful deployment as jurisdictions can impose data retention and audit obligations. These obligations can directly conflict with the privacy guarantees offered by the E2EE platform. Financial regulators like the U.S. Securities and Exchange Commission (SEC) and the Commodity Futures Trading Commission (CFTC) have issued multi-billion dollar fines in the past to firms for failing to retain business communications that were sent using encrypted consumer apps like WhatsApp \cite{sec_offchannel_fines}. Similar challenges are present under the EU's Markets in Financial Instruments Directive (MiFID), which calls for complete and tamper proof communication records for certain transactions \cite{mifid_recordkeeping}.

From an enterprise management perspective, the issues are not any easier. Firstly, E2EE prevents the service providers, and potentially even the organisation's IT department, from accessing message content form compliance review or even forensic investigation. Secondly, the overwhelming use of personal devices and Bring Your Own Device (BYOD) policies further increases the risk of employees bypassing the provided official communication channels completely. This can be done either intentionally in the case of malicious insiders, or completely by accident due to negligence \cite{cisa_insider}. Either way, essential audit trails can be easily lost.

In order to somewhat reconcile compliance and encryption, some enterprise focused E2EE platforms do implement secure message journaling or Data Loss Prevention (DLP) hooks that bascially act as controlled decryption endpoints for authorised compliance officers \cite{aws_wickr_bcc}. While this does help enterprises to comply with regulations, it does also introduce potential points of failure and must be implemented carefully in order to avoid creating backdoors into the system.

\subsubsection{Evolving Threat Landscape}

Quantum computing represents a real risk to widely deployed public-key algorithms as both NIST and the UK's National Cyber Security Centre have urged organisations to move to using post-quantum cryptography (PQC) as soon as possible, well before 2035. This in an attempt to mitigate the "harvest now, decrypt later" strategy that is being leveraged by advanced adversaries \cite{nist_pq_timeline,ncsc_qc_warning}. Surveys reveal that a growing portion of organisations, especially in critical sectors, anticipate quantum decryption capabilities within a decade or sooner \cite{capgemini_qsurvey}. 

Another threat to be considered when it comes to the future threat landscape, are organised crime groups or nation state threat actors, since they have access to large budgets, manpower, as well as new methods such as exploiting AI for targeted surveillance, social engineering, and cryptographic attacks. Although it is still in its early stages, research into AI-augmented encryption with adaptive, context aware cryptographic schemes appears to show promise as a potential solution for future E2EE innovation. This technology would be able to do things such as enabling dynamic parameter tuning based on content sensitivity or threat level \cite{encgpt_dynamic,ai_adaptive_encryption}.

\section{Innovative E2EE Mechanisms and Future Directions}
The challenges facing E2EE have created a new wave of innovation in cryptographic mechanisms and privacy-enhancing technologies. The future of E2EE is not just about strengthening existing protocols but also about creating entirely new mechanisms that address the persistent vulnerabilities of the digital age.

\subsection{Enhancing the "Ends": New Paradigms for Secure Endpoints}
In this subsection we present recent solutions and emerging technologies that directly address the vulnerability of endpoints

\subsubsection{SNE2EE (Secure Node End-to-End Encryption)}

One of the previous attempts to mitigate the endpoint vulnerabilities in E2EE systems was the Secure Node End-to-End Encryption (SNE2EE) framework \cite{velagala2022enhancing,maglaras2022end} which was motivated by the realisation that, while encryption in transit is strong and reliable, the endpoint devices themselves remain the weakest link in the system. In order to overcome the endpoint risks such as spyware, stalkerware, or OS level exploits, SNE2EE proposed the idea of architectural isolation of cryptographic functions from the user-facing device.

The design involves delegating the events such as key exchange, ratchet progression, as well as the actual encryption/decryption to a secondary "secure node" which is physically separate from the primary user device. There were several prototypes that were developed in order to validate this concept, including a secondary mobile phone, a Raspberry Pi, as well as dedicated Bluetooth modules. The primary device which in this case was often a smartphone, would serve only as the interface for writing or viewing a message and then the external secure node carried out the actual cryptographic operations.

The strength of this method lies in its ability to compartmentalize trust so that even if the primary device is compromised by malware, the threat actor cannot directly access session keys or intercept decrypted plaintext since these operations occur only on the secure node. 

\subsubsection{ARSecure (Augmented Reality Secure Messaging)}

Building upon the foundations that were laid by SNE2EE, the ARSecure project sought to address the same class of endpoint vulnerabilities but in a more user friendly and complete package. 
With this idea in mind we created the proof-of-concept project ARSecure which aimed to directly address these somewhat overlooked vulnerabilities by essentially extending the E2EE trust boundary to the creation and display environments of augmented reality (AR) messaging. 

The ARSecure system was developed as a research project by Hamish Alsop \cite{alsop2024arsecure} in order to explore how E2EE principles can be adapted to wearable devices, in this case, smart glasses. To attempt to effectively counter the aforementioned risks, ARSecure uses a dedicated cryptographic module that is held seperate from the general AR runtime, making sure that key management and encryption operations are executed in its own hardened section of the system. The AR interface functions only as a rendering layer, presenting the plaintext to only the intended user, thanks to  the WaveGuide display right next to the wearers eye.

Not only did this solution help to mitigate risks such as over-the-shoulder attacks by having the display right in front of the user's eye, limiting who can see the screen, it also helped to provide a solution to the risk of CSS. This was achieved due to the fact that the glasses are a dedicated device, separate from your everyday mobile phone. By off-boarding the encryption/decryption as well as viewing of messages to a dedicated hardened device rather than your everyday mobile device, which is constantly exposed to all sorts of threats from the internet, the risk of being spied on can be dramatically reduced.

\subsection{Complementary Technological Solutions for Enhanced Endpoint Security}

This subsection presents hardware-based security and Zero Trust Architecture solutions along with more dynamic, software-driven technologies that can further protect endpoints and user data.

\subsubsection{Trusted Execution Environments (TEEs)}

Trusted Execution Environments (TEEs) provide a compelling, hardware enforced method of safeguarding the cryptographic operations from a potentially compromised operating system. It does this by isolating sensitive code and data within a secure "enclave", and it aims to guarantee confidentiality and integrity, even if the host system has malware or is compromised on a root-level \cite{schneider_teedesign, trusted_env_wiki}.

Leading implementations of this include:
\begin{itemize}
  \item \textbf{ARM TrustZone}: This segregates the CPU into "secure" and "normal" worlds, protecting critical operations from OS-level compromise by designating exclusive hardware resources for each domain \cite{zhu_trustzone}.
  \item \textbf{Intel SGX}: This creates encrypted enclaves within memory that remain inaccessible to higher-privilege code. Despite its strong isolation model, SGX does suffer from limited memory, a restrictive programming model, as well as vulnerability to side-channel attacks (e.g., Foreshadow, Plundervolt) \cite{search9, news20, news21}.
  \item \textbf{Apple Secure Enclave}: This integrates a dedicated processor for handling encryption, biometric authentication, and key storage, enabling high-assurance isolation ideally suited for security on mobile endpoints.
\end{itemize}

The upsides of TEEs are clear but they do have their limitations, some are prone to hardware-level exploits such as speculative execution flaws and voltage-induced faults \cite{news20, news21}. 

\subsubsection{Hardware Security Modules (HSMs) and Secure Elements (SEs)}

Hardware Security Modules (HSMs) and Secure Elements (SEs) offer the strongest form of endpoint protection that is currently available. Both of these methods offer tamper resistant hardware environments for key use but they target different contexts and form factors.

HSMs are dedicated cryptographic processors that tend to be rack mounted or external appliances, and they are designed to generate, manage, and store keys within a hardened enclosure. All of the cryptographic operations happen within this device which helps to ensure that secret keys never leave the secure boundary that it creates \cite{yubico_hsm, thales_hsm}. HSMs are also certified to hi assurance standards such as FIPS 140-2/3 or Common Criteria EAL4-EAL7, and are commonly deployed within enterprise environments \cite{wiki_hsm, wiki_ibm4767, wiki_ibm4768}, like IBM's 4746 and 4768 PCIe coprocessors. 

On the other hand, SEs are compact, embedded chips that can be found in smartphones, SIM cards, passports, and much more due to their small size and low power requirements. SEs provide a space for secure storage as well as isolated execution for applications that are running within a controlled environment, separate from the device's main OS. They are often certified to GlobalPlatform standards EAL4+, which provides strong protection for sensitive data like cryptographic keys or even biometric data \cite{tropicsquare_comparison, wiki_secure_element}.

\begin{table}[h!]
\centering
\caption{Comparison of HSMs and Secure Elements}
\begin{tabularx}{\linewidth}{|X|X|X|}
\hline
\textbf{Characteristic} & \textbf{HSMs} & \textbf{SEs} \\
\hline
Form Factor & Rack-mounted appliance or external device & Embedded chip in mobile/IoT devices \\
\hline
Security Level & Highest (FIPS 140-2 Level 4+, EAL7) & Strong (EAL4+, tamper-resistant) \\
\hline
Performance & High throughput, support for complex cryptography & Limited compute; optimized for key storage and simple operations \\
\hline
Flexibility & Custom logic, high crypto agility & Limited to built-in applets, firmware restrictions \\
\hline
Use Cases & PKI, banking, centralized infrastructure & Mobile payments, identity, E2EE endpoint protection \\
\hline
\end{tabularx}
\end{table}

\subsubsection{Confidential Computing}

Confidential computing builds upon the principles of the trusted execution environments (TEEs) discussed previously, from local devices to cloud and distributed infrastructure. It makes sure that data remains encrypted while being actively processed by applications in memory. \cite{confidential_computing_cons2022}.

Practical implementations of this can be seen in things like Intel Trust Domain Extensions (TDX), or AMD's Secure Encrypted Virtualisation (SEV), or Apple's Private Cloud Compute \cite{intel_tdx_whitepaper,amd_sev_snp,apple_privatecloud2023}. 

From a E2EE messaging standpoint, these types of technologies could allow privacy preserving services like encrypted search, or AI spam detection, or even content moderation to be performed on messages without actually exposing the paintext to the cloud providers. 

\subsubsection{ Hardware Wallets / Security Keys}

Using physical devices such as hardware wallets and FIDO2 security keys can provide tamper resistant storage as well as on-device cryptographic operations, both of which can offer strong defence against endpoint compromise in E2EE messaging. Some devices use secure elements and confirmation interfaces to isolate the key operations, examples of this can be seen in Ledger and Trezor, while others can offer non-exportable signing and key exchange that can be enforced with user presence (e.g. touch or biometrics), with examples of this being seen in Yubikey's solution as well as Google Titan Keys \cite{ledger_cspn2022,trezor_learn,fido_specs_overview,ctap2_v22,yubico_tech_manual,google_titan_key}. 

In the context of E2EE messaging, these types of devices could be used to hold long term identity keys to ensure private material never leaves the secure hardware. If hardware keys are implemented well, then it can really raise the bar for attackers without creating too much friction in terms of user experience. Making use of compact form factors, such as NFC fobs or smart wearables, can be beneficial for maintaining usability while still reaping the security benefits of the external hardware device. \cite{google_app_overview,wired_titan_passkeys,soups2019_reese_tokens}. 

\subsection{Advancements in Core Cryptographic Primitives}
This subsection details the foundational technologies that are evolving to strengthen E2EE from a cryptographic perspective.

\subsubsection{Post-Quantum Cryptography (PQC)}

As the world of quantum computing advances, traditional public-key methods, such as RSA or ECC, face increasing threats due to algorithms like Shor's, which can break them quickly once sufficiently powerful quantum computers exist. This then gives rise to the concern of the "store now, decrypt later" threat, where adversaries capture encrypted data in the present day in order to eventually decrypt it once quantum computing matures \cite{csc_store_now,sslstore_hndl,nist_pqc_faq}.

In order to try and combat this threat, a new class of algorithm based on lattice, hash, or code problems, is being increasingly standardised by bodies like NIST. These methods, which are designed to be able to resist both classical and quantum attacks, are being implemented in products such as Zoom's post-quantum E2EE, or Apple's iMessage \cite{zoom_pq_e2ee,news_apple_pq3}.

Even though PQC algorithms tend to have larger keys as well as higher computational costs, benchmarks indicate that some implementations, such as Kyber or Dilithium, can in fact be practical on mobile hardware, although payload size and CPU usage are nontrivial when compared to classical methods \cite{mdpi_mobile_pqc}. 

\subsubsection{Homomorphic Encryption (HE)}

HE allows for computation on cipher-texts without actually having access to the pain-text or secret keys, which can enable privacy preserving operations even within untrusted environments, \cite{wired_he_definition, wiki_he_def} which is very useful for E2EE systems. This essentially allows E2EE data to be processed by servers or AI for analytics or spam filtering, without every decrypting or compromising the data.

\begin{table}[H]
\centering
\scalebox{0.8}{%
\begin{tabular}{|p{3cm}|p{7cm}|}
\hline
\textbf{Type} & \textbf{Description / Characteristics} \\
\hline
Partially Homomorphic & Supports a single type of operation (e.g., addition or multiplication) on ciphertexts \cite{wiki_he_def}. \\
\hline
Somewhat Homomorphic & Supports limited operations, typically both addition and multiplication, but only for a bounded number of computations \cite{wiki_he_def}. \\
\hline
Fully Homomorphic (FHE) & Supports arbitrary computations on encrypted data. Originally impractically slow with Gentry’s 2009 design being up to a trillion times slower than plaintext computation, but modern improvements have reduced this to roughly a million times slower, which still remains a major barrier for mobile or real-time applications \cite{wired_he_definition}. \\
\hline
\end{tabular}
}
\caption{Types of Homomorphic Encryption and Their Performance Characteristics}
\end{table}

Recent advances have managed to improve the practicality of this technology, for example the CKKS method supports approximate arithmetic, which is ideal for machine learning on encrypted data \cite{wiki_he_def}. This can be seen in libraries like Microsoft SEAL and OpenFHE as they implement these methods to support efficient batching and bootstrapping for encrypted computation \cite{ms_seal_lib, openfhe_lib}. 

There are also hybrid HE methods emerging such as GuardML" which blends both symmetric and homeomorphic encryption in order to enable efficient encrypted neural network inference while keeping low overhead on mobile devices \cite{guardml_paper}. Alternatively there is other work such as "THE-X" which has enabled transformer model inference on encrypted data by using approximation methods that are tuned for HE while managing to keep performance acceptable \cite{thex_paper}. 

\subsubsection{Secure Multi-Party Computation (MPC)}

Secure Multi-Party Computation (MPC) essentially allows several parties to jointly work on computing a function over private inputs without revealing them. This method effectively replaces the role of a trusted third party while also aligning with decentralisation principles in E2EE. 

This works with "secret sharing" or "garbled circuits", where basically inputs are split up into random shares that are then split up across the participants. This means that no single participant can revoke the secret alone, but together they can compute the desired output \cite{yao1982protocols, goldreich1987mcp}. 

Some applications of this technology include Private Set Intersection (PSI) for contact discovery \cite{pinkas2018psi}, collaborative fraud detection across institutions \cite{lindell2020secure}, as well as decentralised key management using distributed key shares \cite{araki2016high}.  

Although there are modern protocols such as SPDZ which have managed to reduce costs via preprocessing and efficient secret sharing \cite{damgaard2012multiparty}, unfortunately, MPC still suffers from performance and communication overheads, which is somewhat limiting its deployment in latency-sensitive mobile E2EE. 

\subsubsection{Verifiable Computation / Zero-Knowledge Proofs (ZKPs)}

Zero Knowledge Proofs (ZKPs) are essentially cryptographic protocols that are used to allow one party, typically called the prover, show to the other party, typically called the verifier, that they know something to be true but without actually telling the verifier the thing that they know \cite{goldwasser1985knowledge}. This can make ZKPs a powerful tool for use cases where sensitive data must remain hidden, but trust in computation or identity must be established. For example, in E2EE applications, they can enable a user to prove possession of valid credentials without actually revealing the credentials themselves, which is more or less perfect for privacy-preserving authentication \cite{ben2019zkapps}.

Then there is Verifiable Computation, which does tend to rely on ZKPs, that extends on the previously mentioned concepts by allowing users to outsource computations to other untrusted servers while still retaining the assurance that the results are correct \cite{gennaro2013non}. This type of method can significantly strengthen cloud-backed E2EE systems, as it enables servers to assist with message routing or spam filtering while ensuring correctness, even if the servers are not fully trusted.

There are have been some recent advancements in this area to make ZKPs more efficient as well as scalable, some examples of this include zk-SNARKs (Succinct Non-Interactive Zero-Knowledge Proofs) or zk-STARKs (Scalable Transparent Arguments of Knowledge) \cite{ben2014succinct, ben2018stark}. zk-SNARKs are able to provide compact proofs that are suitable for blockchain or identity verification, provided that they have a trusted setup, but on the other hand, zk-STARKS avoid the need for a trusted setup and offer post-quantum security, but have the drawback of larger proof sizes. 

\subsubsection{Oblivious RAM (ORAM)}

While encryption is able to protect the actual data contents, the metadata, such as what addresses are accessed and how frequently they are accessed, can leak sensitive information; this is where Oblivious RAM (ORAM) can offer a solution. ORAM is a cryptographic technique that is designed to conceal memory access patterns from an adversary that is observing a system's interactions with storage \cite{goldreich1996oram}. It does this by continually shuffling and re-encrypting data blocks to ensure that the sequence of memory requests essentially looks the same as random noise to any would be observer \cite{stefanov2013oram}. 

This can be useful in situations where threat actors may be monitoring cloud servers or compromised endpoints. Since ORAM hides access patterns, it can prevent the attacker from inferring any user behavior, even if the ciphertext is exposed.

Unfortunately the practicality of ORAM is limited by its computational overhead as well as its bandwidth overhead. The older ORAM constructions can impose a logarithmic overhead per access, but some more advanced methods like Path ORAM or Ring ORAM manage to reduce costs but still remain quite resource-intensive \cite{stefanov2013oram,ren2015oram}. However, current research is exploring hardware-assisted ORAM and hybrid models that combine ORAM with TEEs to reduce costs \cite{ren2015oram}.

\begin{table*}[htbp]
\centering
\scriptsize
\renewcommand{\arraystretch}{1.25}
\setlength{\tabcolsep}{5pt}
\begin{tabular}{p{1.5cm} p{3cm} p{3cm} p{3cm}}
\toprule
\textbf{Work} & \textbf{Approach} & \textbf{Main Contribution} & \textbf{Key Limitation} \\
\midrule
Folkerts et al.~\cite{folkerts2025testing} & Split-model with encrypted attention head & Reveals vulnerability in IP protection & Susceptible to model extraction attacks \\
Kumar et al.~\cite{kumar2025tfhe} & Compiler-integrated TFHE code generation & First benchmark for TFHE code with LLM optimization & Standard LLMs struggle without RAG/few-shot prompting \\
Rho et al.~\cite{rho2024encryption} & HE-friendly transformer design & 6.94$\times$ faster fine-tuning, 2.3$\times$ faster inference & Computational overhead remains significant \\
De et al.~\cite{de2024privacy} & GPU-accelerated FHE for GPT-2 & First GPU extension to OpenFHE; efficient encrypted inference & Approximation of activations may affect accuracy \\
Bae et al.~\cite{bae2025privacy} & Hybrid cloud–local framework & Privacy-preserving CoT reasoning with encrypted search & Relies on cooperation between local and cloud models \\
\bottomrule
\end{tabular}
\caption{Comparison of recent approaches integrating end-to-end encryption with LLMs.}
\label{tab:e2ee_llms}
\end{table*}

\section{Integrating End-to-End Encryption with Large Language Models}

End-to-end encryption (E2EE) and fully homomorphic encryption (FHE) are emerging as critical techniques for safeguarding sensitive queries in large language model (LLM) deployments. Recent works have explored different integration strategies, ranging from compiler-assisted code generation to hybrid local cloud inference pipelines. This section summarizes key contributions in this area. Table~\ref{tab:e2ee_llms} summarizes the key contributions of the surveyed works.

\subsection{Split-Model Security Risks}
Folkerts et al. \cite{folkerts2025testing} propose an investigation into the security of split-model approaches that combine fully homomorphic encryption (FHE) with large language models (LLMs) to enable privacy-preserving inference while protecting model intellectual property (IP). In current deployments, users send encrypted data to a server that executes an encrypted attention head layer and then returns the results for local continuation, ensuring both user privacy and partial model confidentiality. However, the study reveals a critical vulnerability: this single-layer split model setup is susceptible to attacks that allow users to recover the server’s proprietary neural network components, effectively bypassing intended IP protections. Through detailed analysis, the work demonstrates the practicality of such extraction attacks and highlights the limitations of relying solely on encrypted computation for safeguarding model assets. The authors conclude by discussing possible mitigation strategies, underscoring the need for stronger protections in privacy-preserving LLM deployments.

\subsection{Compiler-Assisted TFHE Code Generation}
Kumar et al. \cite{kumar2025tfhe} propose a compiler-integrated framework that leverages large language models (LLMs) and agentic optimization techniques to generate Fully Homomorphic Encryption (FHE) code over the torus (TFHE), with a focus on logic gates and ReLU activation functions. While TFHE offers strong potential for privacy-preserving applications such as secure machine learning, multi-party computation, and private blockchain, its complexity and lack of accessible tools hinder broader adoption. To address this, the study evaluates open- and closed-source LLMs on their ability to produce TFHE code, analyzing compilability, structural fidelity, and error rates. Findings show that standard LLMs face significant limitations, but enhancements like retrieval-augmented generation (RAG) and few-shot prompting substantially improve accuracy and reliability. By establishing the first benchmark for TFHE code generation, this work demonstrates how LLMs, when coupled with domain-specific guidance, can help bridge the gap between cryptographic expertise and practical usability in FHE systems.

\subsection{HE-Compatible Transformer Architectures}
Rho et al. \cite{rho2024encryption} propose a privacy-preserving framework that adapts transformer architectures for compatibility with homomorphic encryption (HE), aiming to address the privacy risks associated with personalized large language model (LLM) interactions. While HE enables secure computation on encrypted data, its integration with LLMs is hindered by the heavy computational demands of transformer models. To overcome this challenge, the study introduces an HE-friendly transformer design optimized for inference after private fine-tuning. By leveraging LoRA-based fine-tuning and Gaussian kernel techniques, the approach achieves notable efficiency gains—6.94× speedup in fine-tuning and 2.3× in inference—without sacrificing performance compared to unencrypted models. The results demonstrate a practical proof of concept for deploying privacy-preserving LLM services in sensitive domains where data protection is essential, with the proposed implementation made openly available on GitHub.

\subsection{GPU-Accelerated FHE for LLM Inference}
De et al. \cite{de2024privacy} propose a fully encrypted framework for running large language models (LLMs) using fully homomorphic encryption (FHE), aiming to protect sensitive queries processed on third-party cloud platforms. Since client inputs may include confidential data such as intellectual property or medical records, outsourcing LLM inference raises significant privacy risks. To overcome the challenges of inefficient FHE implementations and incompatibility with LLM activation functions, the study introduces two key contributions. First, it delivers the first open-source GPU-accelerated FHE implementation as an extension to the popular OpenFHE library, achieving substantial speedups in core operations such as bootstrapping. Second, it develops and evaluates approximation techniques for LLM activation functions to preserve model accuracy while improving efficiency. Applied to GPT-2, the proposed system achieves forward-pass times significantly faster than CPU baselines, with only minimal loss in accuracy and perplexity when tested on HellaSwag, LAMBADA, and ARC benchmarks. This work demonstrates the feasibility of practical, privacy-preserving LLM inference through encrypted computation.

\subsection{Hybrid Privacy-Preserving Frameworks}
Bae et al. \cite{bae2025privacy} propose a hybrid privacy-preserving framework that enables large language models (LLMs) to act as personal agents while safeguarding sensitive user data such as emails, calendars, and medical records. Existing approaches force a trade-off between relying on powerful but untrusted cloud-based LLMs or weaker local models. To bridge this gap, the study introduces Socratic Chain-of-Thought Reasoning, where a non-private query is first processed by a powerful external LLM to generate a structured chain-of-thought and sub-queries. These sub-queries are then embedded and matched against a user’s private data using a homomorphically encrypted vector database capable of sub-second semantic search over one million entries. The retrieved documents, once decrypted, are combined with the chain-of-thought prompt and processed by a smaller trusted local model to produce the final output. Evaluations on the LoCoMo long-context QA benchmark show that this cooperative setup—pairing GPT-4o with a local Llama-3.2-1B—achieves performance gains of up to 7.1 percentage points compared to GPT-4o alone. This work establishes a promising direction for splitting reasoning tasks across strong untrusted LLMs and weaker local ones to enhance both functionality and privacy.

\section{Conclusions}

End to End Encryption remains a cornerstone of modern-day secure digital communication, yet it is neither a panacea, nor is it a static solution. The effectiveness of E2EE depends on much more than just how strong the mathematics of the encryption algorithms are, it is also heavily reliant on the security of the endpoints, the integrity of the implementation, as well as the awareness and the behaviour of the users.

The current and future research into areas like post-quantum cryptography, homeomorphic encryption, and even innovative endpoint-centric solutions such as ARSecure, show us that meaningful progress is being made towards more resilient, as well as user friendly systems for E2EE.

At the same time, there is also the broader societal and regulatory landscape to think about when looking at the resilience of encryption. The reactions to new legislation and legal mandates clearly illustrate the tensions between lawful access and the privacy of users, as more often governments are pushing for Client Side Scanning, or Backdoors into messaging platforms. These kinds of pressures make it very clear that the future of E2EE will not just be shaped by innovation within the field, but also by political, legal, and ethical choices.

In the end, sustaining strong encryption requires a combination of the development of strong encryption platforms, as well as the cultivation of informed and responsible use of these platforms. E2EE is not optional, it is essential in ensuring user's right to privacy, yet its future is far from guaranteed. Without vigilance from policy makers, technologists, and even society as large, regulatory pressure does risk eroding the protection and privacy that E2EE offers. This means that safeguarding strong encryption is not just a technological imperative, but it is also a societal responsibility.

\bibliographystyle{vancouver-modified}
\bibliography{sample-vancouver}

\end{document}